I. Title:

Large-area deposition of protective (Ti,Al)N coatings onto polycarbonate

II. Authors:

Lena Patterer[1]*, Sabrina Kollmann[2]*, Teresa de los Arcos[2], Leonie Jende[1], Soheil Karimi Aghda[1], Damian M. Holzapfel[1], Sameer Aman Salman[1], Stanislav Mráz[1], Guido Grundmeier[2], Jochen M. Schneider[1]

[1] Materials Chemistry, RWTH Aachen University, Kopernikusstr. 10, 52074 Aachen, Germany

[2] Technical Chemistry, Universität Paderborn, Warburger Str. 100, 33098 Paderborn, Germany

-------------------------

* Corresponding authors:   patterer@mch.rwth-aachen.de
                           sschwid2@mail.upb.de

**Abstract**

Polycarbonate (PC) and protective (Ti,Al)N coatings exhibit extremely different material properties, specifically crystal structure, thermal stability, elastic and plastic behavior as well as thermal expansion coefficients. These differences present formidable challenges for the deposition process development as low-temperature synthesis routes have to be explored to avoid a thermal overload of the polymer substrate. Here, a large-area sputtering process is developed to address the challenges by systematically adjusting target peak power density and duty cycle. Adhering (Ti,Al)N coatings with a critical residual tensile stress of 2.2 ± 0.2 GPa are obtained in the pulsed direct current magnetron sputtering range, whereas depositions at higher target peak power densities, realized by high power pulsed magnetron sputtering, lead to stress-induced adhesive and/or cohesive failure. The stress-optimized (Ti,Al)N coatings deposited onto PC with a target peak power density of 0.036 kW cm$^{-2}$ and a duty cycle of 5.3% were investigated by cross-cut test confirming adhesion. By investigating the bond formation at the PC | (Ti,Al)N interface, mostly interfacial $CN_x$ bonds and a small fraction of (C-O)-(Ti,Al) bonds are identified by X-ray photoelectron spectroscopy, indicating reactions at the hydrocarbon and the carbonate groups during deposition. Nanoindentation reveals an elastic modulus of 296 ± 18 GPa for the (Ti,Al)N coating, while a Ti-Al-O layer is formed during electrochemical impedance spectroscopy in a borate buffer solution, indicating protective passivation. This work demonstrates that the challenge posed by the extremely different material properties at the interface of soft polymer substrates and hard coatings can be addressed by systematical variation of the pulsing parameters to reduce the residual film stress.

## 1. Introduction

Poly(bisphenol A carbonate), generally called polycarbonate (PC), is a widely used thermoplastic material characterized by high impact resistance, ductility, flame retardancy, and relatively low production costs leading to the production of 6.1 million tons in 2020 [1,2]. Typical applications for PC and its blends are, for instance, optical storage devices, automotive and aircraft components, as well as construction panels [2,3]. However, its susceptibility to stress corrosion cracking upon moisture exposure, low solvent resistance, and relative softness limit its application [3]. Nevertheless, by functionalizing the surface of PC with a protective coating, its application can be widened to more demanding conditions, like humid atmospheres [4] and surface wear-intensive applications [4–6].

(Ti,Al)N (space group $Fm\bar{3}m$, B1 NaCl prototype structure) is commonly used as a protective hard coating for cutting applications due to its high thermal stability [7], outstanding oxidation [8], as well as wear resistance [9]. Especially, the former two properties qualify (Ti,Al)N for application as a protective coating for polymers.

High power pulsed magnetron sputtering (HPPMS) is commonly used for the synthesis of (Ti,Al)N and has the advantage of synthesizing dense coatings, even in the absence of substrate heating [10], which is for most polymers a decisive factor due to their low melting temperature ($T_g$ (PC) ~ 150 °C [11]). In this context, the application of a substrate bias potential can be often omitted for HPPMS as well [10], since large ion energies can lead to high residual stress in the coating [12,13] promoting delamination from the substrate.

HPPMS is characterized by a high power applied in several tens of µs long pulses at the target resulting in a high fraction of ionized target and gas species, which contribute to a dense coating formation due to the intense low-energy ion irradiation [14,15]. Moreover, the high target peak power causes a short nucleation period and, hence, a high island density of the growing film on the substrate [10]. For TiN, it was observed that the deposition rate during the on-time is 100 times larger compared to the direct current magnetron sputtering (DCMS) process with the same time-averaged power [10]. Especially for polymers, this might be an important factor since, for Au depositions onto a trimethylcyclohexane bisphenol polycarbonate (TMC-PC), it was reported that the interface formation significantly depends on the deposition rate [16]. It was observed that a high deposition rate results in a sharp substrate | coating interface formation, whereas diffusion of Au atoms into the polymer substrate was detected under low deposition rate conditions, leading to chemically weak interfacial interaction and an unstable interface upon heating [16]. This data suggests that the short nucleation time during HPPMS may enable strong and stable interfacial bond formation, and hence, sufficient adhesion of the coating on the substrate.

Besides interfacial diffusion, other influencing factors have to be taken into account for the deposition onto soft polymer substrates compared to rigid substrates, like steel or Si. The coefficient of thermal expansion (CTE) of polymers is very high compared to ceramic coatings (e.g. $CTE_{PC}$ ~ 8 × $CTE_{(Ti,Al)N}$ [11,17]) leading to thermal stress at the interface upon cooling [18]. Additionally, extrinsic stress contributions resulting from interactions of polymer substrates with the environment have to be considered for PC, specifically, the influence of post-deposition moisture uptake [19]. Bradley *et al.* [19] observed a significant increase in tensile stress in thin films grown on PC within the first hours after deposition upon atmosphere exposure and a return to equilibrium state afterwards, which can be rationalized by the moisture uptake and subsequent water vapor diffusion into the substrate. Hence, it is important to design the protective coating for PC in a way that it withstands this range of stress evolution upon atmospheric moisture exposure without cracking. Furthermore, the mechanical properties of substrate and coating have to be considered for load-bearing applications, as the mismatch of elastic moduli at the interface might lead to buckling of the coating [20] or crack initiation [21].

So far, no literature exists that systematically analyzes the interaction between (Ti,Al)N coatings deposited onto PC. Nevertheless, studies of isostructural $TMN_x$ (TM = Ti, Cr) coatings deposited onto various polymers [22–26], identified parameters affecting interface and coating quality.

One of these parameters is the deposition time, defining both the film thickness and the substrate temperature. Chaiwong *et al.* [22] obtained adhering 80 nm-thick TiN coatings onto PC (substrate temperature $T_S$ up to 45 °C, deposition time ~ 40 min), whereas cracking and buckling were observed as the thickness was increased to 120 nm ($T_S$ up to 55 °C, deposition time ~ 60 min). Moreover, the film density is crucial as shown by Zhang *et al.* [25] who compared CrN coatings deposited onto acrylonitrile butadiene styrene (ABS) by either DCMS or HPPMS, demonstrating that HPPMS leads to a denser microstructure, superior corrosion resistance, a higher elastic modulus ($E_{DCMS}$ ~ 240 GPa, $E_{HPPMS}$ ~ 270 GPa) and a delayed crack initiation during thermal alternation tests. By comparing Ti and TiN coatings deposited onto PC (thickness < 100 nm), a better adhesion under load was observed for TiN coatings since at loads of 100 mN, crack formation for the Ti coating was observed [27]. Nevertheless, in many studies, Ti interlayers are used to improve the adhesion between TiN coatings and polymer substrates [22–24] since the lower elastic modulus mismatch at the substrate | coating interface causes a lower interfacial stress concentration under load [28]. In another study, the influence of N concentration on the surface roughness of $CrN_x$ coatings deposited onto ABS was studied by Pedrosa *et al.* [26] reporting a increased surface roughness of pure Cr with elongated granular morphology ($R_a$ ~ 42 nm) compared to $Cr_{0.75}N_{0.25}$ forming pyramid-like column tops ($R_a$ ~ 25 nm).

Overall, the challenge of joining interfaces that consist of two materials with extremely different properties, such as ceramic (Ti,Al)N and polymeric PC, motivated the here communicated research. A systematic variation of pulsing conditions for a large-area (Ti,Al)N deposition onto PC was carried out, followed by an extensive investigation of the microstructural, mechanical, and oxidation properties of the coatings.

## 2. Experimental methods

Most PC substrates were prepared by spin-coating (2000 rpm) a 5 wt.% PC solution, containing PC pellets (additive free, product # 4315139, Sigma Aldrich) dissolved in tetrahydrofuran (anhydrous, 99.9% purity, inhibitor-free, Sigma Aldrich), on Si(001) substrates (1 × 1 cm$^2$, Crystal GmbH). For the atomic force microscopy (AFM), the electrochemical oxidation, and the cross-cut test, a filtered solution of 2.5 wt.% PC in 1,4 dioxane (99.8% purity, Sigma Aldrich) was used for the substrate preparation. Furthermore, nanoindentation and additional reference measurements were carried out on a (Ti,Al)N coating grown directly onto a Si(001) substrate (1 × 1 cm$^2$, Crystal GmbH) to obtain more realistic mechanical properties of the coating by excluding the elasto-plastic behavior of the polymer substrate.

The depositions were performed in an industrial-scale CemeCon CC 800/9 deposition system with one rectangular composite $Ti_{0.5}Al_{0.5}$ target (99.7% purity, Plansee Composite Materials GmbH) with the dimensions of 8.8 × 50 cm$^2$. The target-to-substrate distance was 8 cm, and the substrate holder was kept at a floating potential during depositions. No intentional heating was applied to the substrate, however, the temperature change during deposition was monitored by a thermocouple mounted to the substrate holder. The base pressure before deposition was always < 0.5 mPa. For the (Ti,Al)N depositions, the total gas pressure during deposition was ~ 0.4 Pa, while the flow ratio of Ar:N$_2$ was set to 10:1. Metallic TiAl interlayers were deposited in a pure

Ar atmosphere with a gas pressure of ~ 0.36 Pa. During depositions, a digital oscilloscope was used to track the peak current and target peak power density at the target. For the systematic variation study of the pulsing parameters using a Melec SIPP2000USB-10-500-S pulser, the deposition time was 10 min (10+10 min for the bilayer systems), while for the subsequent characterization study, (Ti,Al)N coatings were synthesized with a deposition time of 120 min to obtain a coating thickness of ~ 1 µm (**Figure 1**). For the HPPMS and pulsed DCMS depositions, the target peak power density was varied between 0.035 – 0.62 kW cm$^{-2}$ by changing the applied time-averaged power at the target, while the duty cycle was varied between 1–5.3% by keeping the pulse on-time ($t_{ON}$) constant at 50 µs and varying the pulse off-time ($t_{OFF}$) between 893-4950 µs.

For the interfacial bond analysis, a deposition time of 4 s (nominal thickness < 1 nm, **Figure 1**) was used to detect the interface with X-ray photoelectron spectroscopy (XPS), which typically reaches an information depth of < 10 nm [29]. To explore the effect of nitrogen on the interfacial bond formation by XPS, a PC substrate was treated with an N$_2$ plasma by applying a pulsed DC voltage to the substrate holder (power density = 2.5 W cm$^{-2}$, frequency = 250 kHz, $t_{off}$ = 1616 ns) in a pure N$_2$ atmosphere ($p_{N2}$ = 0.72 Pa) for 4 s (**Figure 1**). An overview of the prepared samples for the different analysis methods is shown schematically in **Figure 1**.

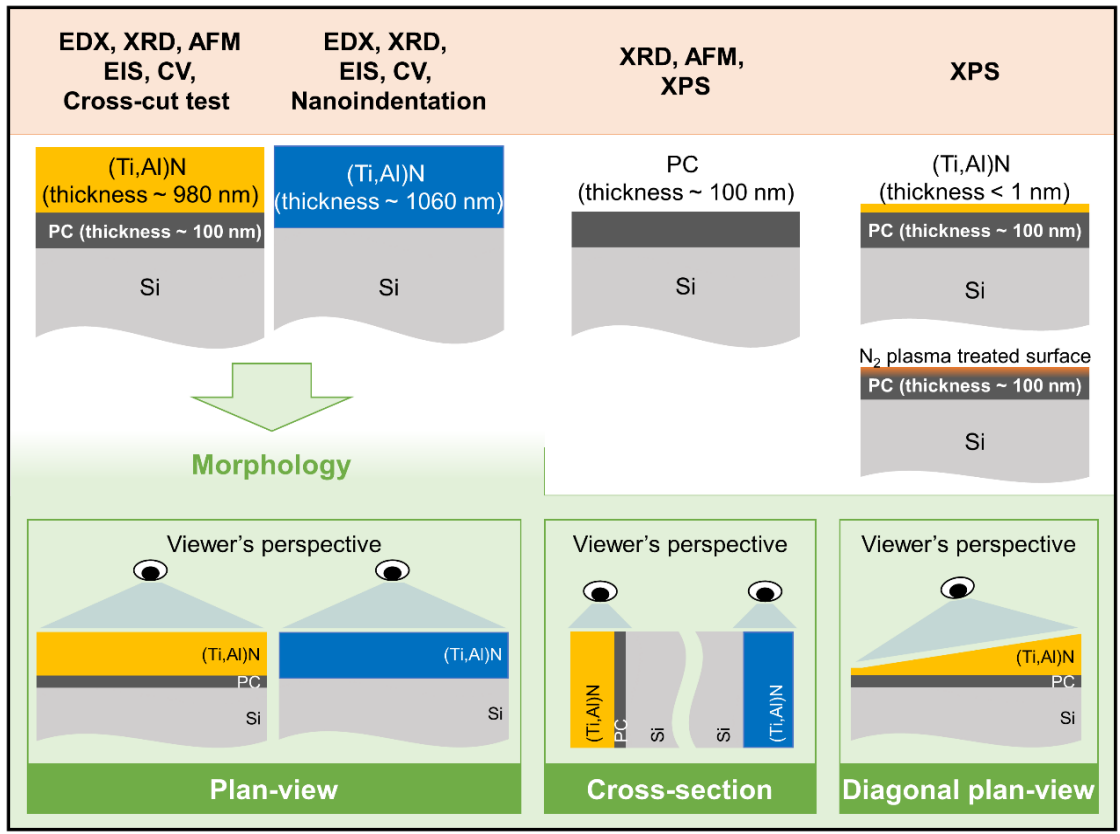

**Figure 1.** Schematic of prepared (Ti,Al)N samples on different substrates for the various characterization methods.

The structural and stress analyses were carried out in Bruker AXS D8 Discover General Area Detection Diffraction System with a Cu K$_\alpha$ source. An acceleration voltage of 40 kV and a current of 40 mA were used for X-ray diffraction (XRD). The measured diffractograms show the detected intensity at *2θ* angles in the range of 19-81°.

For the chemical analysis of the deposited coatings, a JEOL JSM-6480 scanning electron microscope with an EDAX Genesis 2000 device was used by applying an acceleration voltage of 10 kV. A Ti$_{0.269}$Al$_{0.210}$O$_{0.004}$N$_{0.517}$ coating [30], measured by elastic recoil detection analysis (ERDA), served as a standard for the energy-dispersive X-ray spectroscopy (EDX) measurements.

The microstructure of the coatings was characterized by scanning (transmission) electron microscopy (S(T)EM). Plan-view and cross-sectional lamellae of the coatings (see **Figure 1**) were prepared by the focused ion beam (FIB) technique using an FEI Helios Nanolab 660[TM] (Hillsboro, OR, USA) equipped with a field-emission microscope. In addition, a diagonal lamella from the plan view of the coating was prepared to attain thickness-induced morphological characteristics. The schematic for the prepared diagonal lamella is shown in **Figure 1**. $Ga^+$ ions were used at the acceleration voltage of 30 kV for the extraction of the lamellae. First, a 1.5 µm-thick Pt layer was grown as protection on the region of interest with a current of 80 pA. This was followed by trench milling at 2.5 nA current and the extracted lamellae were thinned to a thickness of < 100 nm. The same microscope was used to obtain the SEM surface images and STEM micrographs from the prepared lamellae.

The atomic force microscopy (AFM) images were taken with a Dimension ICON from Bruker in tapping mode with a SCM-PIT-V2 cantilever (75 kHz, 3 N m$^{-1}$) under ambient conditions. Measurements were done at three different positions with a scan area of 5 x 5 µm$^2$. The evaluation of the AFM images and determination of roughness $R_{max}$ were done with the NanoScope analysis software.

XPS measurements were performed in a Kratos AXIS SUPRA instrument (Kratos Analytical Ltd.), using monochromatic Al-K$_α$ X-ray radiation ($hv$ = 1486.6 eV) and a hemispherical analyzer. The measurement spot size was 700 µm × 300 µm. Survey scans with a pass energy of 160 eV and a step size of 0.25 eV were carried out (5 sweeps, dwell time of 100 ms), while high-resolution N 1s and C 1s spectra were obtained using a pass energy of 20 eV and a step size of 0.05 eV (20 sweeps, dwell time of 100 ms). For the samples containing a PC signal, a charge neutralizer (low-energy, electron-only source) was applied to avoid charging effects and the binding energy (BE) scale was referenced with respect to the hydrocarbon group of PC at 284.6 eV [31]. For the conducting 1 µm-thick (Ti,Al)N sample, charge neutralization was not required and the BE scale was referenced with respect to the Fermi edge [29].

Quantitative analysis was carried out by using the CasaXPS software package (Casa Software Ltd.), subtracting a Shirley background [32], and applying the manufacturer's sensitivity factors [33]. For the fitting of the high-resolution spectra, a mixed Gaussian-Lorentzian (70%-30%) peak shape was applied, while the full width at half maximum of the C 1s components was constrained to ≤ 1.5 eV according to [31].

To probe the adhesion properties of the (Ti,Al)N coating, a cross-cut test was carried out according to the norm DIN 2409. For this, a standardized cutting tool with parallel arranged blades was used. The distance between the blades was 1 mm. Afterwards, the sample was cleaned in an air flow and characterized with a light microscope. Subsequently, an adhesive tape (Tesa) was applied to the sample and then detached perpendicular to the surface. The sample was then analyzed with the light microscope again.

The elastic modulus and the nanoindentation hardness were determined by nanoindentation in a Hysitron Triboindenter using a Berkovich-geometry diamond tip (radius = 100 nm). The tip area function was calibrated with a fused silica standard. For the analyzed sample, 100 indents with a maximum load of 2 mN were carried out resulting in a contact depth of < 10% with respect to the film thickness of the (Ti,Al)N coating. The reduced modulus was determined from the unloading part of the load-displacement curve following the method of Oliver and Pharr [34] and the elastic modulus was calculated by assuming a Poisson's ratio of $v$ = 0.214 [35].

For the electrochemical impedance spectroscopy (EIS) and cyclic voltammetry (CV), a Reference 3000 potentiostat (Gamry Instruments) was used. The experiments were carried out in borate buffer solution (0.2 M $H_3BO_3$ (for analysis), 0.05 M $Na_2SO_4$ (99%) and 0.05 M $Na_2B_4O_7 \cdot 10\ H_2O$ (99.5%) are all from Sigma-Aldrich) with a pH of 8.3 under ambient conditions using a three-electrode set-up. The reference electrode was an Ag/AgCl electrode, and a gold wire was used as the counter electrode. The EIS measurements were performed with a potential amplitude of ΔU = ± 10 mV at open circuit potential. For the CV studies, 20 cycles with a scan rate of 200 mV s$^{-1}$ were measured in a potential range between -2 and 2 V (SHE).

## 3. Results

### a. Deposition process development

To develop a large-area deposition process, the parameter space of duty cycle vs. nominal target (peak) power density was systematically screened for adhering (Ti,Al)N coatings deposited onto PC. The deposition approaches are summarized in a diagram that was compiled based on Gudmundsson *et al.* [36] (see **Figure 2**): (Ti,Al)N depositions (HPPMS and pulsed DCMS) are shown as squares, TiAl depositions (HPPMS and DCMS) are shown as circles, and TiAl (DCMS) | (Ti,Al)N (HPPMS)-bilayer depositions are shown as triangles. The color code in **Figure 2** indicates the condition of the coating observed immediately after the deposition (green = adhesion, yellow = surface defects, red = coating failure). Selected pictures of the coatings taken immediately after deposition can be found in the supporting information (**Figure S1** and **Figure S2**).

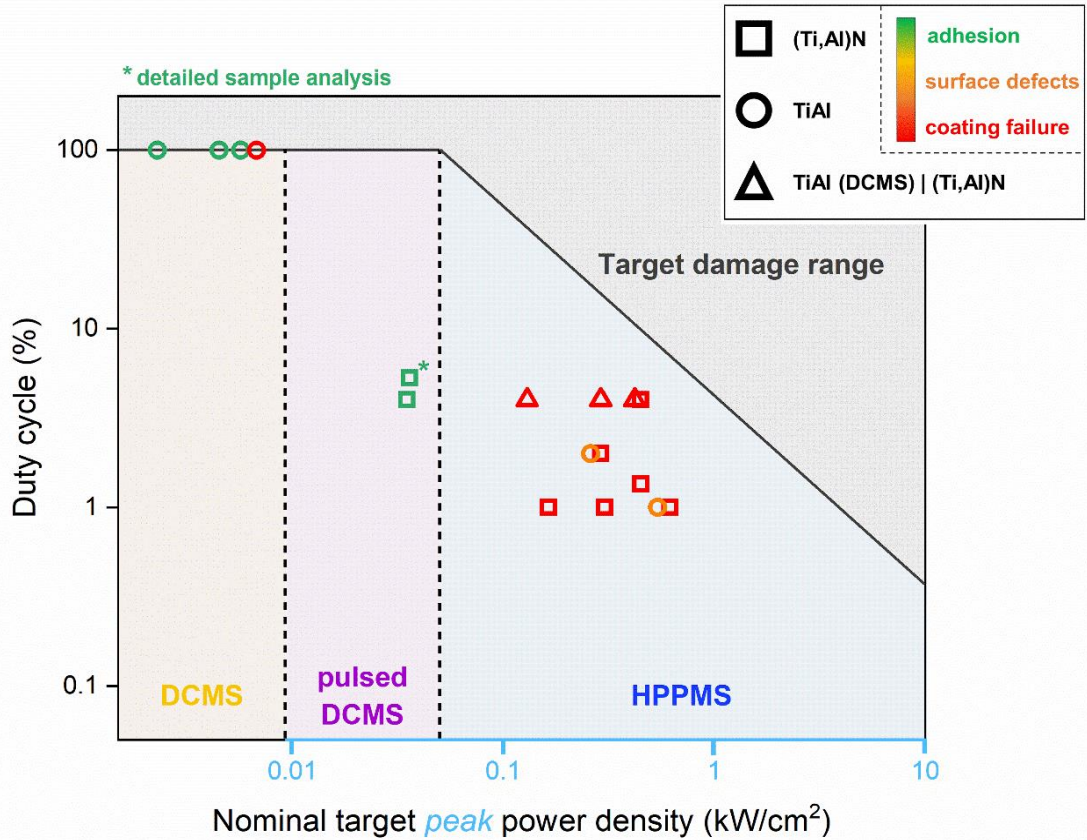

**Figure 2.** (Ti,Al)N, TiAl, and TiAl | (Ti,Al)N-bilayer depositions onto PC with systematic variations of duty cycle and target (peak) power density categorized based on [36]. The color code of symbols (green, yellow, red) indicates the coating condition after deposition.

Since a denser microstructure can be obtained under HPPMS conditions compared to DCMS [10], the first attempts were made to tune duty cycle and target peak power density in the HPPMS range. However, all these depositions resulted in cohesive or adhesive failure of the (Ti,Al)N coatings (red squares in **Figure 2**), mainly caused by cracking and blister formation (supporting information, **Figure S1**).

Due to the lower elastic moduli mismatch of TiAl compared to (Ti,Al)N at the substrate interface (DFT-based $\Delta E_{PC|TiAl}$ < 200 GPa [37] vs. $\Delta E_{PC|(Ti,Al)N}$ > 350 GPa [38]), the synthesis route *via* a metallic TiAl interlayer was tested. While HPPMS TiAl coatings showed some surface defects (orange circles, **Figure 2**), TiAl deposited by DCMS (duty

cycle = 100%) showed good adhesion and a smooth surface (supporting information **Figure S2**) when the power density was kept ≤ 0.006 kW cm$^{-2}$ (green circles, **Figure 2**). A power density of 0.007 kW cm$^{-2}$, however, caused distinct buckling [22] of the TiAl coating (supporting information, **Figure S2**) possibly caused by the substrate heating up to 70 °C.

Following these insights, several experiments with a smooth DCMS TiAl interlayer (power density = 0.002 kW cm$^{-2}$, substrate heating < 45 °C) followed by HPPMS (Ti,Al)N depositions were carried out (red triangles, **Figure 2**). However, none of these TiAl | (Ti,Al)N-bilayer approaches showed satisfying adhesion, instead crack formation was observed, possibly due to stress generation, (supporting information, **Figure S2**). Since the HPPMS conditions cause a high-stress level in the TiAl and (Ti,Al)N coatings at which adhesion onto PC could not be achieved (red and yellow symbols in HPPMS range, **Figure 2**), two (Ti,Al)N depositions (without metallic interlayer) in the pulsed DCMS mode were performed, resulting in successful depositions of homogenous, adhesive coatings (green squares, **Figure 2**). Since the (Ti,Al)N coating deposited at a slightly higher duty cycle (marked as □* in **Figure 2**) showed better results regarding the aimed chemical atomic composition of Ti$_{0.25}$Al$_{0.25}$N$_{0.5}$, these conditions (duty cycle = 5.3%, target peak power density = 0.036 kW/cm$^2$) were selected to synthesize samples for the further characterization in this study. Deposition details, besides duty cycle and target peak power density, for this PC | (Ti,Al)N sample can be found in **Table 1**. The homogeneous surface of the coating without any signs of cracks or buckling is shown in the low-magnification SEM image in **Figure 3**.

**Table 1.** Deposition parameters for the PC | (Ti,Al)N sample marked as □* in **Figure 2**

| Target | Ti$_{0.5}$Al$_{0.5}$ (8.8 × 50 cm$^2$) |
|---|---|
| Time-averaged power (W) | 400 |
| Target peak power density (kW/cm$^2$) | 0.036 |
| Duty cycle (%) | 5.3 |
| $I_{peak}$ (A) | 31 |
| $t_{OFF}/t_{ON}$ | 50 µs/893 µs |
| $p_{Ar}$ (mPa) | 360 |
| $p_{N2}$ (mPa) | 35 |
| $p_{base}$ (mPa) | < 0.5 |
| Deposition time (min) | 120 |
| Deposition rate (nm/min) | 8.2 |
| Substrate temperature (°C) | < 42 (no intentional heating) |
| Target-to-substrate distance (cm) | 8 |

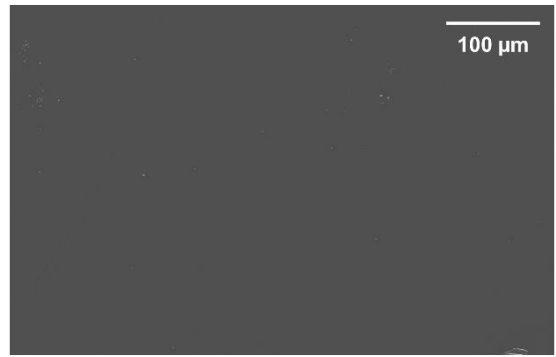

**Figure 3**: Large-area, low-magnification SEM surface image of the adhering (Ti,Al)N coating onto PC (duty cycle = 5.3%, target peak power density = 0.036 kW cm$^{-2}$).

### b. Chemical and structural analysis

**Table 2** shows the chemical composition of the (Ti,Al)N coatings deposited onto PC as well as onto Si substrates analyzed by EDX. The detected 4-6 at.% O can be rationalized by the partly porous microstructure of the coating, shown in chapter **3c** below, enabling O diffusion from the surface along the column boundaries [39]. Additionally, the rather low deposition rate of 8.2 nm min$^{-1}$ (onto PC) promotes the incorporation of O originating from residual water in a vacuum chamber [40,41]. In this way, the slightly higher deposition rate of (Ti,Al)N on Si (8.8 nm min$^{-1}$) might explain the lower oxygen concentration. Due to the similar results, the substrate seems to play a minor role in the chemical composition of the here deposited (Ti,Al)N coatings.

**Table 2.** Chemical composition of (Ti,Al)N coatings measured by EDX (calibrated with an ERDA measured (Ti,Al)N coating [30])

|  | Ti (at.%) | Al (at.%) | N (at.%) | O (at.%) |
|---|---|---|---|---|
| PC \| (Ti,Al)N | 22 | 25 | 47 | 6 |
| Si \| (Ti,Al)N | 22 | 26 | 48 | 4 |

To analyze the crystal structure, XRD was performed on pristine PC, PC | (Ti,Al)N, and Si | (Ti,Al)N samples (see **Figure 4**). Since PC is an amorphous material, the final slope of the amorphous hump between 19 and 30° [42] is visible for the pristine PC as well as for the PC | (Ti,Al)N sample. Independent of the substrate, the (Ti,Al)N coatings show a cubic structure with a preferred (111) orientation, while a smaller signal of the (220) orientation is evident, which is slightly more pronounced for the PC | (Ti,Al)N sample. However, no (200) peak at ~ 43° [43] is observed. In B1 NaCl structures, the growth along the [111] direction exhibits the densest atom column arrangement, while the [001] growth direction is the most open channeling direction and exhibits the lowest

surface energy ($E_{DFT}$ = 1.34 J m$^{-2}$ [44]) [43]. The (220) plane ($E_{DFT}$ = 3.39 J m$^{-2}$) exhibits a slightly lower surface energy compared to the (111) plane ($E_{DFT}$ = 4.20 J m$^{-2}$) [44]. Hence, the (111) preferred orientation is expected for low-energy growth, while the (200) texture is preferred under conditions with higher adatom mobility (high temperature, high ion energy) [43]. Generally, the ion kinetic energy ($E_k$) of species arriving at the substrate can be calculated by the following formula:

$$E_k = E_i + qe\,(V_{pl} - V_s)$$

where $E_i$ is the initial ion kinetic energy before entering the substrate sheath, $q$ is the ion charge state, $e$ is the elementary charge, $V_{pl}$ is the plasma potential, and $V_s$ is the substrate bias potential. In this study, the floating potential at the substrate holder was $V_s \sim$ -20 V as previously measured for the here utilized experimental setup [12,13]. Based on previous plasma diagnostics for HPPMS (Ti,Al)N depositions ($V_{pl} \sim$ 5 eV, $E_i \sim$ 0 eV [12]), a kinetic energy of $E_k \sim$ 25 eV of singly-charged ions ($q$ = 1) can be expected for our (Ti,Al)N depositions.

Following the argumentation above, the estimated ion kinetic energy in our depositions matches the low-energy deposition conditions causing a (111) preferred orientation as previously observed for Ti$_{0.5}$Al$_{0.5}$N [43]. Different templating effects of PC and Si (with top layer SiO$_x$) substrates [45], and as well as the ~ 80 nm difference in coating thickness might cause the minor differences in the detected (220) intensity relative to the (111) signal.

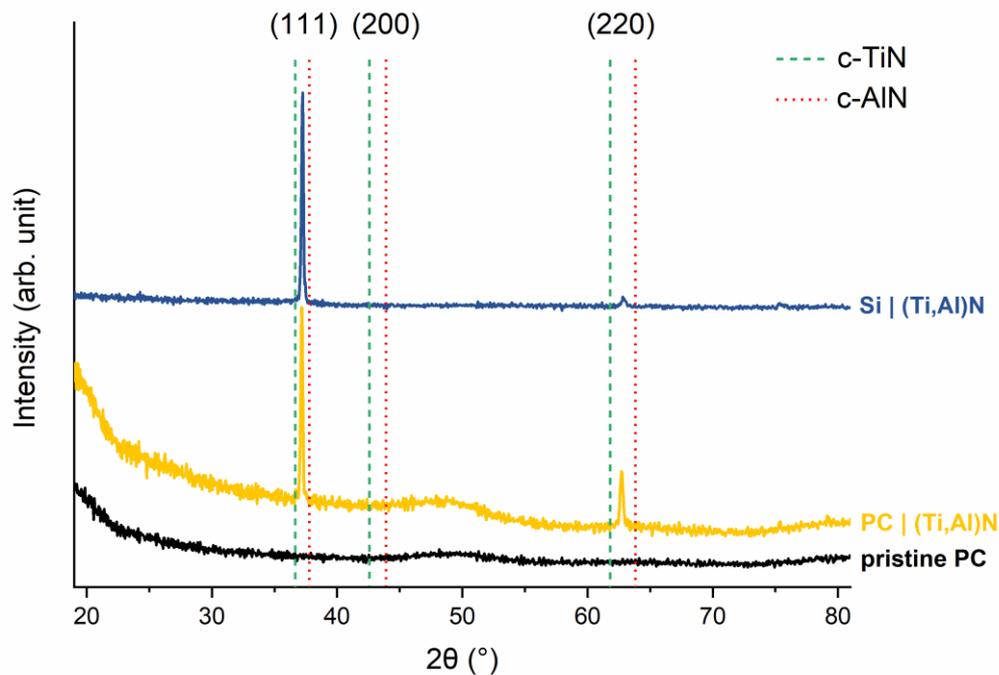

**Figure 4.** Diffractograms of pristine PC (black), PC | (Ti,Al)N (yellow), and Si | (Ti,Al)N (blue).

### c. Morphology

First, the cross-section and surface morphology of the (Ti,Al)N coatings, both grown onto PC and Si, were examined (**Figure 5** a-d), revealing, independent of the substrate, a coarse-columnar microstructure with the presence of porosities along the elongated columns.

The surface-near region of the cross-sections indicates a dome-shaped surface topography (**Figure 5** a & c) that is typical for transition metal nitrides during competitive columnar growth observed for the transition zone (zone T) of the structure zone model (SZM) [45–47]. Additionally, the plan-view SEM micrographs from the surface of both coatings, shown in **Figure 5** b & d, reveal a pyramidal or faceted surface morphology as it was also observed for low-energy growth of ScN [48] and CrN [49]. For CVD-deposited (Ti,Al)N, the pyramidal surface morphology was explained by the preferential growth along the [111] direction forming the tip of the pyramid, while the three pyramid facets constitute the {001} planes [50].

The cross-sections of the coatings can be divided into two regions: the substrate-near microstructure includes finer grains with fewer pores, while after ~ 100 nm, larger V-shaped columns with increasing porosity can be identified. According to Barna and Adamik [46], zone T of the SZM is characterized by a growth accompanied by some surface diffusion but a strongly limited grain boundary diffusion. This results in a grain structure that is not homogenous throughout the film thickness, as observable in **Figure 5** a & c. The resulting V-shaped columns have a kinetically favored orientation - here (111) - and overgrow the kinetically disadvantaged grains [45]. The competitive growth of these kinetically favored columns causes also atomic shadowing which promotes the formation of pores [45] (see arrows, **Figure 5** a & c).

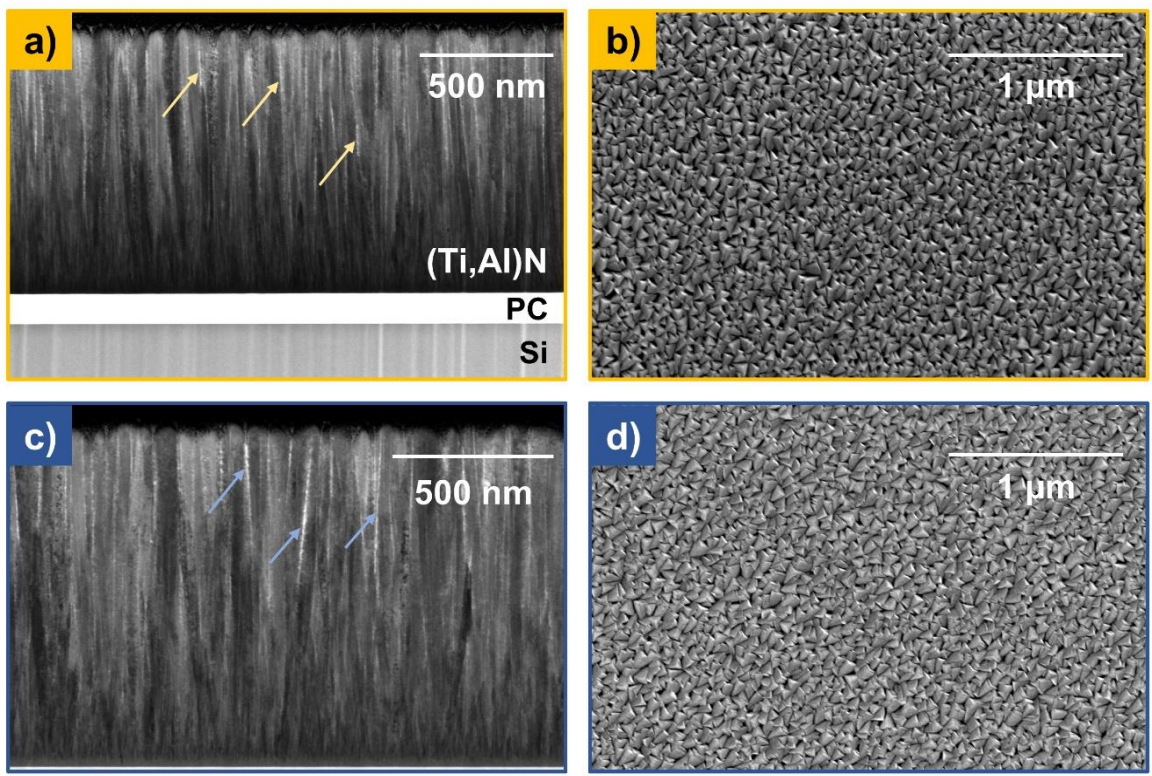

**Figure 5.** a) and c) cross-section STEM and b) and d) surface SEM morphology of (Ti,Al)N coatings deposited onto a) and b) PC, as well as c) and d) Si, respectively. Arrows indicate the presence of pores.

The plan view along the coating's diagonal (**Figure 6**) enables the analysis of the microstructural evaluation as a function of the thickness in further detail. Again, it is clearly visible that the grain size in the substrate-near region is much finer and coarsens as the coating becomes thicker. Almost no pores are observed in the substrate-near region (< 1% pore area, see quantification in supporting information **Figure S3** a), while more and larger pores, shown as white spots, can be observed in the surface-near region (~ 5% pore area in **Figure S3** b). Even though a porous microstructure is observed, the requirements for a protective coating can be still fulfilled due to the dense structure at the substrate interface.

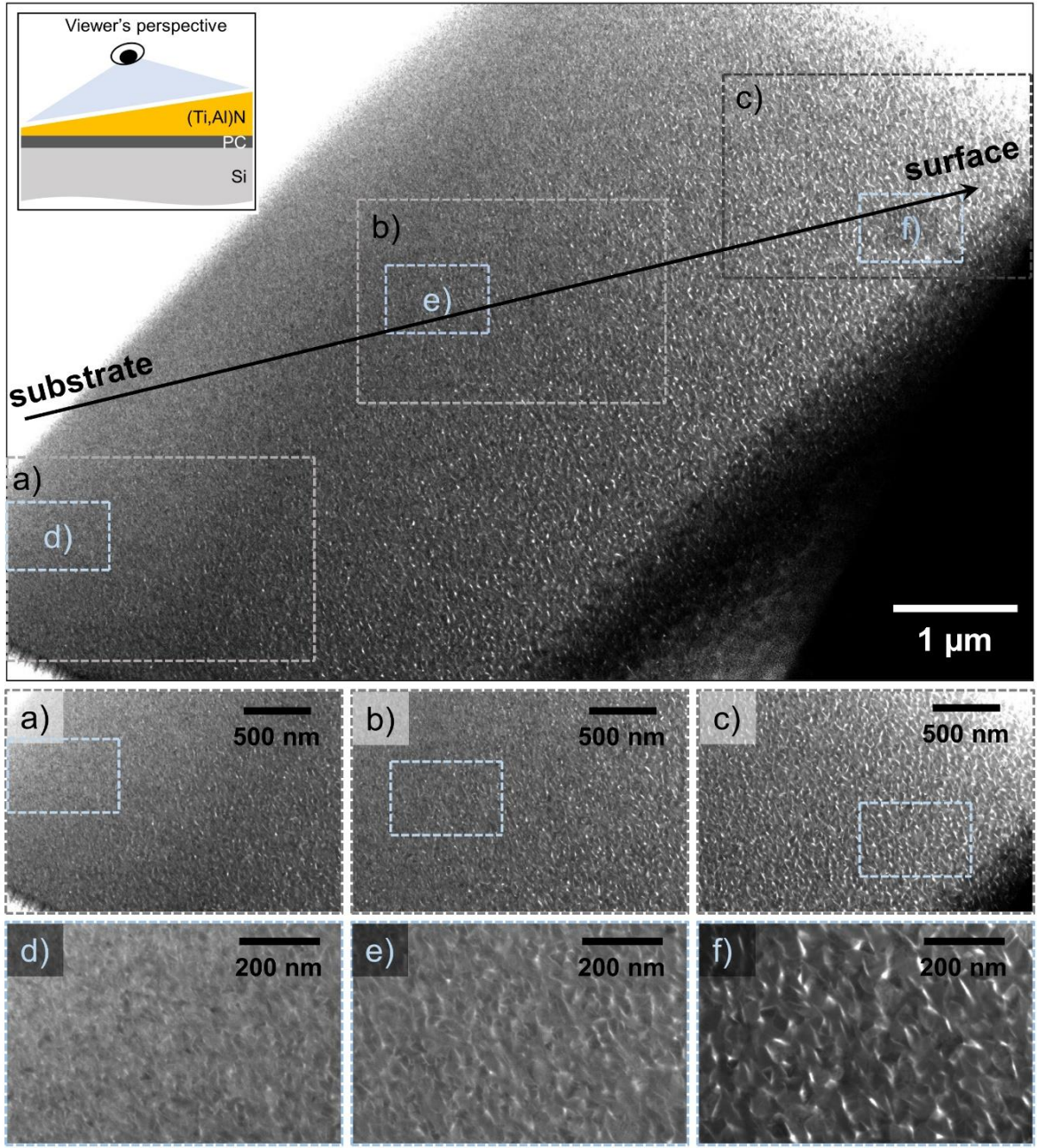

**Figure 6.** Bright-field STEM images show morphology at a diagonal plan view of the PC | (Ti,Al)N sample with different magnifications.

### d. Surface and interface roughness

In addition to the morphology, the surface roughness of the spin-coated PC and the as-deposited PC | (Ti,Al)N sample were analyzed by means of AFM (see **Figure 7** a & b). Moreover, the interface between the (Ti,Al)N coating and the PC is analyzed by dissolving the polymer phase. In this way the interface of the (Ti,Al)N coating could be measured (**Figure 7** c).

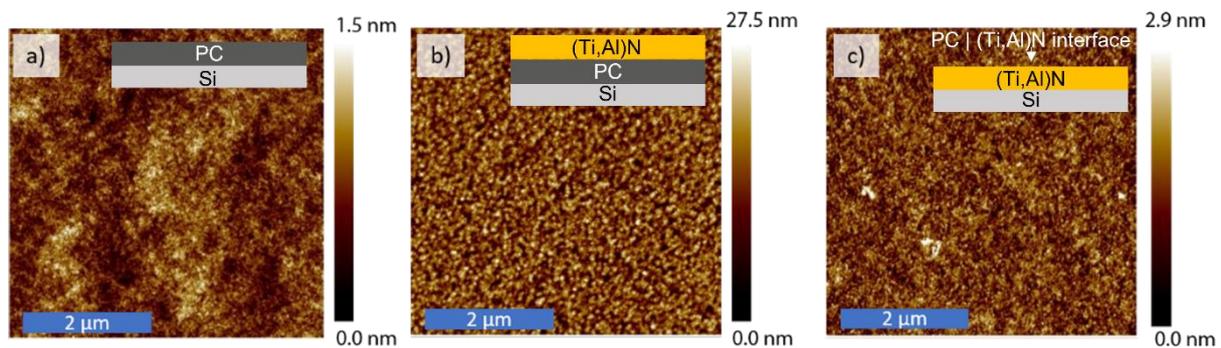

**Figure 7.** AFM images a) spin-coated PC surface, b) surface of (Ti,Al)N coating as deposited onto PC, and c) interface between PC and TiAlN

The average roughness of the spin-coated PC was measured to be 3.7 ± 2.2 nm. After the deposition of the (Ti,Al)N coating, the roughness increased to a mean value of 33.8 ± 12.9 nm. The morphology of the coating surface as studied by AFM agrees well with the observation made by SEM (see **Figure 6**). The AFM image of the interface between the (Ti,Al)N coating and PC after dissolving the PC phase is illustrated in **Figure 7** c. The roughness of this opened (Ti,Al)N interface is 4.6 ± 0.1 nm (rms value) and thus by a factor of 10 smaller compared to the outer (Ti,Al)N surface. The morphology of the interface resembles that of the PC surface which hints at an almost defect-free (Ti,Al)N | PC interface. As the interfacial roughness is slightly larger than the initial PC surface roughness, we assume that during deposition interfacial atomic diffusion takes place which might contribute to the adhesion of the coating.

### e. Interfacial bond analysis

To investigate the bond formation at the PC | (Ti,Al)N interface, an XPS analysis of a (Ti,Al)N thin film (thickness < 1 nm) deposited onto PC was carried out. The N 1s and the C 1s signals were analyzed by comparing the interface spectra with reference spectra of pristine PC, a PC surface that was treated with an $N_2$ plasma, and a 1 µm-thick (Ti,Al)N coating. The N 1s spectrum of the (Ti,Al)N reference sample (**Figure 8** I. a) contains the (Ti,Al)(O,N) component at a BE of 396.7 eV and a satellite signal at BE of 398.2 eV which is in good agreement with literature [51]. The reference sample of the $N_2$-plasma treated PC (**Figure 8** I. c) shows two N 1s components at BEs of 399.3 and 400.7 eV, matching the reported BE values of single- and triple-bonded $CN_x$ groups (N-C, N≡C), as well as double-bonded $CN_x$ groups (N=C), respectively [52]. When the N 1s spectrum of the PC | (Ti,Al)N interface is considered (**Figure 8** I. b), two components can be distinguished: The component with the higher intensity at BE = 399.8 eV can be attributed to interfacial $CN_x$ bonds, while the smaller (Ti,Al)(O,N) component at BE = 396.6 eV originates from the growing (Ti,Al)N coating covering the interface (compare **Figure 8** I. a).

For the analysis of the interfacial C 1s spectrum (**Figure 8** II.), the reference samples of pristine PC and $N_2$-plasma treated PC are considered first (a & c). It is evident that the component at BE ~ 286.0 eV is increasing due to $N_2$ plasma interaction (see **Figure 8** II. a & c), which can be rationalized by the formation of C-N bonds (BE = 285.9 eV [52]) overlapping with the $C_{ring}$-O component at BE = 286.1 eV [31]. Additionally, another component appears at a higher BE = 287.7 eV, which can be assigned to C=N bonds [52]. Both $CN_x$ bond contributions are also evident for the interfacial C 1s spectrum of PC | (Ti,Al)N (marked cyan in **Figure 8** II. b). However, another component becomes evident at BE = 289.0 eV for the C 1s spectrum of

PC | (Ti,Al)N (marked purple in **Figure 8** II. b), which can be ascribed to (C-O)-(Ti,Al) bonds, indicating reactions between the metal atoms and the carbonate group of PC [53]. However, no evidence of C-Ti bonds at BE ~ 282.0 eV is observed as it was detected for the metallic PC | Ti interface [53].

Overall, the XPS analysis reveals that the interface formation between (Ti,Al)N and PC is mainly defined by $CN_x$ bonds and a small contribution of (C-O)-(Ti,Al) bonds.

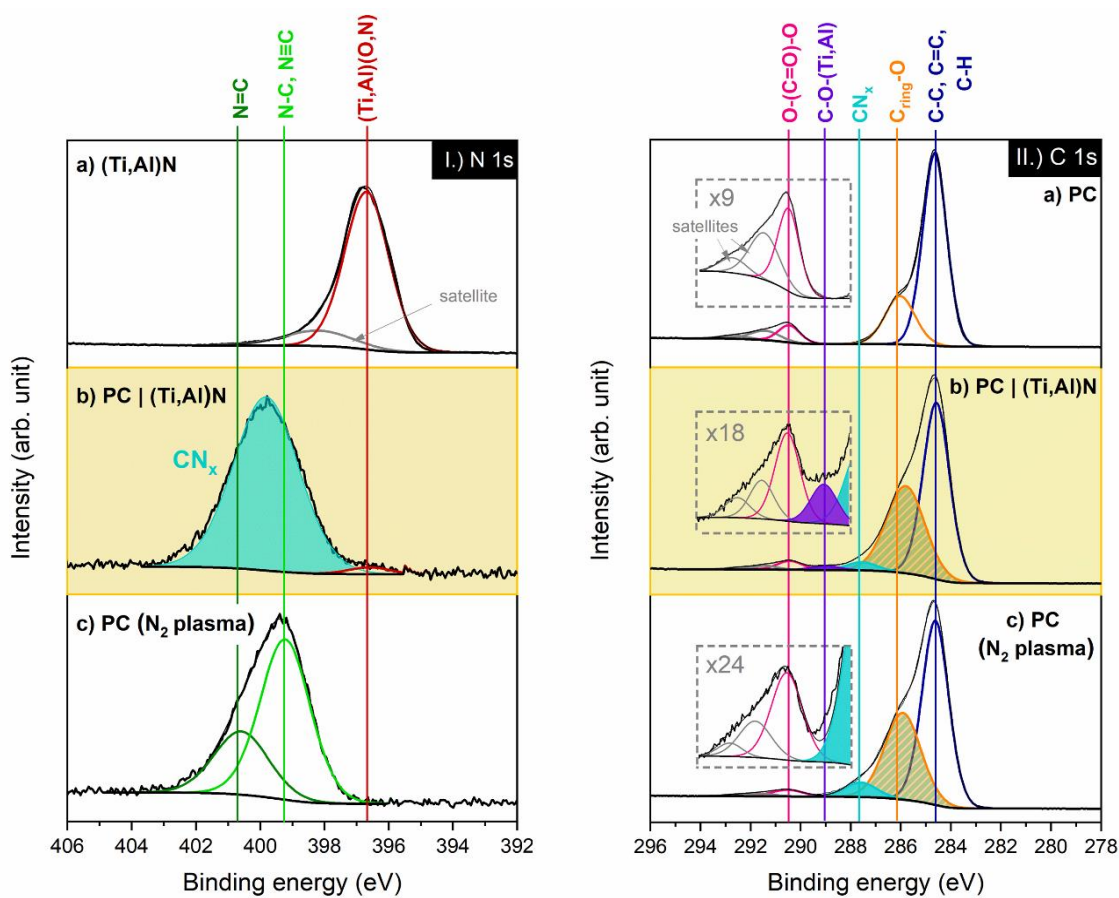

**Figure 8.** XPS analysis of I.) the N 1s signal of a) a 1 µm-thick (Ti,Al)N coating, b) the PC | (Ti,Al)N interface, and c) PC treated with an $N_2$ plasma, as well as
II.) the C 1s signal for a) pristine PC, b) the PC | (Ti,Al)N interface, and c) PC treated with a $N_2$ plasma (insets are magnification of the spectra between BE = 288-294 eV).

### f. Stress analysis

For both (Ti,Al)N samples deposited onto PC and Si, an X-ray stress analysis using the $\sin^2(\Phi)$ method [18] was carried out and the stress values are shown in **Figure 9** (yellow and blue areas). For the PC | (Ti,Al)N sample a tensile stress of 2.2 ± 0.2 GPa is determined, while the Si | (Ti,Al)N samples exhibits a lower value of 1.7 ± 0.2 GPa. The measured deviation of the (Ti,Al)N coating deposited either on PC or Si is about ~ 20% (**Figure 9**). However, it can be assumed that the total stress difference (Δσ, **Figure 9**) is even higher due to the fact that the coefficient of thermal expansion (CTE) of PC is significantly higher compared to (Ti,Al)N ($CTE_{PC}$ ~ 8 × $CTE_{(Ti,Al)N}$ [11,17]), leading to a (negative) compressive stress at the interface (yellow striped area in **Figure 9**). In contrast, Si has a smaller CTE ($CTE_{Si}$ ~ 0.38 × $CTE_{(Ti,Al)N}$ [54]) leading to a low (positive) tensile stress at the interface (blue striped area, **Figure 9**). In addition, it is reasonable to assume that the dense microstructure in the substrate-near region (**Figure 6** d) contributes towards a compressive stress state in the first ~ 100 nm of the coating independent of the substrate. Then, overall, the measured tensile stress is an integral value of a not homogeneously distributed stress values throughout the coating thicknesses which vary from the interface to the film surface. Considering the morphology of the coating, it is expected that the high tensile stress mainly originates from the underdense microstructure in the upper ~ 800 nm defined by the growth conditions of zone 1/T from the SZM [47], which seems to result in an overall higher tensile stress state for the PC | TiAlN sample.

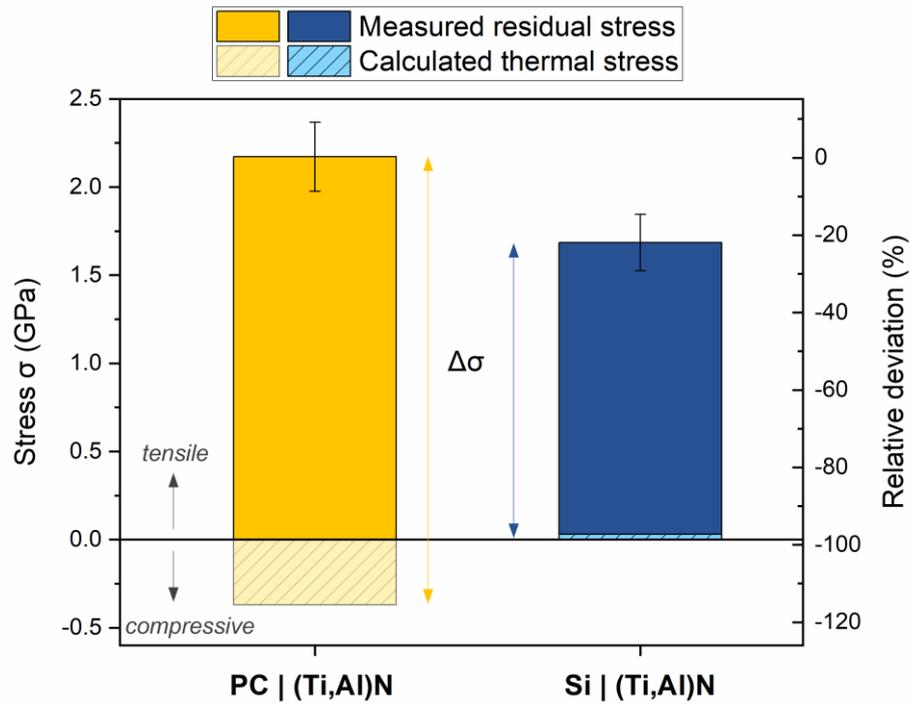

**Figure 9.** Stress state of PC | (Ti,Al)N and Si | (Ti,Al)N coatings was determined by the XRD-sin$^2$(Φ) method [18]. Thermal stress was calculated by using the CTE of the interface materials [55].

g. **Qualitative adhesion testing and mechanical properties**

The adhesion of the deposited (Ti,Al)N coating on spin-coated PC was characterized by the cross-cut test. Additionally, an adhesive strip was pulled-off from the surface to analyze the adhesion after the cross-cut test. **Figure 10** presents the corresponding microscope images before and after pulling-off the adhesive strip from the surface.

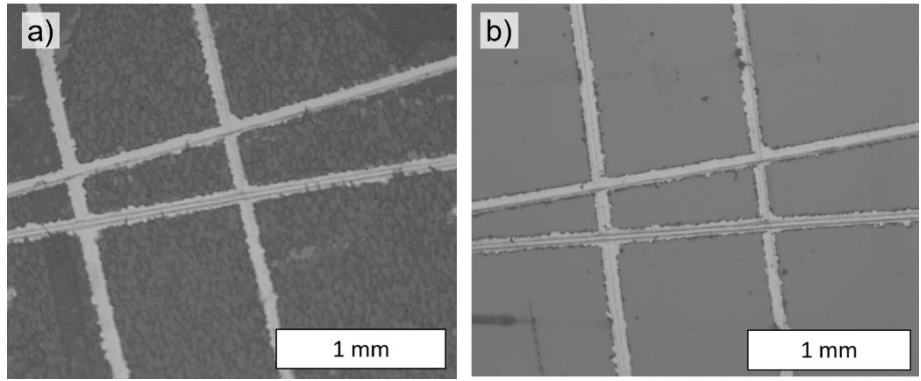

**Figure 10.** Cross-cut test of the PC | (Ti,Al)N sample a) before, and b) after removing the adhesive strip.

The cross-cut test before pulling-off the adhesive stripe (**Figure 10** a) indicates only a light detachment of the (Ti,Al)N coating on the edges where the cuts cross. The evaluation of these measurements was done according to the standard ISO 2409:2013(E). The result shows a replacement below 5%. The additional pull-off of an adhesive strip leads to no change in the detachment of the as-deposited (Ti,Al)N from the spin-coated PC substrate. Thus, the coating also adheres after this test.

The very good adhesion of the (Ti,Al)N coating on the spin-coated PC could be explained by nanoscopic mechanical interlocking and chemical bonding between the hard coating and polymer film.

The nanoindentation measurements to determine the elastic modulus and the hardness were performed on the (Ti,Al)N that was deposited onto Si since the observed substrate effect of the soft polymer was too large for the PC | (Ti,Al)N sample. By performing nanoindentation, an elastic modulus $E$ of ~ 296 ± 18 GPa and a hardness of 13 ± 1 GPa was determined for the (Ti,Al)N coating deposited onto Si. For (Ti,Al)N deposited onto PC, a higher tensile stress was measured (**Figure 9**) which decreases the elastic modulus [17]. Based on previous DFT calculations for (Ti,Al)N, a tensile stress increase of 1 GPa reduces the elastic modulus by only 11.4 GPa [17]. In this way, the stress difference between the PC | (Ti,Al)N and the Si | (Ti,Al)N

samples (**Figure 9**) would result in a $\Delta E$ within the error bar of the experimentally determined $E_{Si\,|\,(Ti,Al)N}$ of ~ 296 ± 18 GPa.

Previously, DFT calculations performed on a defect-free $Ti_{0.25}Al_{0.25}N_{0.5}$ cell determined a theoretical elastic modulus of ~ 390 GPa at 0 K [38]. The lower value measured here can be explained by the residual tensile stress in the coating [56], and the partly porous microstructure [57]. As previously observed, the measured 4-6 at.% O content might also contribute to the reduction in the elastic modulus of (Ti,Al)N [58]. Nevertheless, when comparing the mechanical properties of the (Ti,Al)N coating with the mechanical properties of PC (< 3 GPa [59]), a significant improvement is evident. Hence, (Ti,Al)N can be considered a suitable wear protection for PC.

### h. Electrochemical stability

The (Ti,Al)N coating deposited onto PC was also analyzed concerning the electric behavior with cyclic voltammetry (CV) and electrochemical impedance spectroscopy (EIS) in a borate buffer solution (pH 8.3). The results are compared with measurements on (Ti,Al)N deposited onto Si. This allows to characterize the influence of the substrate on the electrochemical stability. The CV curves are presented in **Figure 11**.

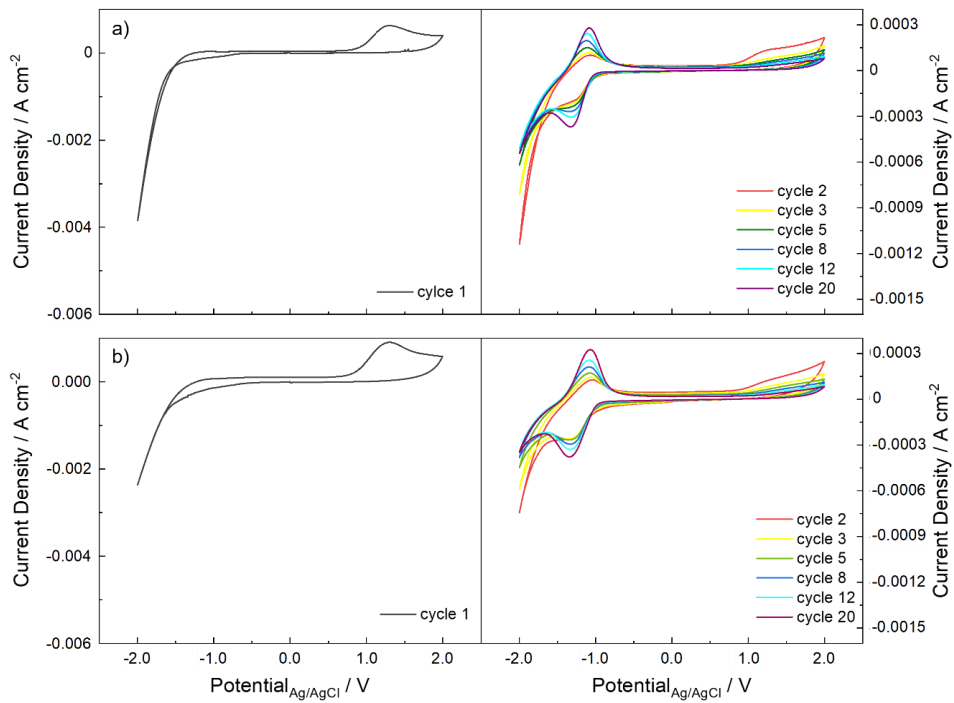

**Figure 11.** Polarization curves in borate buffer (pH 8.3) of a) Si | (Ti,Al)N, and b) PC | (Ti,Al)N

The electrochemical measurements of the PC | TiAlN sample (contact via the surface of the film) indicate that the thin films lead to an electric conductivity of the surface. In the first cycle, an irreversible peak appears at a potential of approximately 1.3 $V_{Ag/AgCl}$ for both substrates, indicating oxidation of the (Ti,Al)N surface. In the following cycles, a reversible redox peak in the range between -1.0 and -1.5 $V_{Ag/AgCl}$ is observed. This originates from the redox pair $Ti^{3+}/Ti^{4+}$ and illustrates oxidation of the surface leading to a mixed ultra-thin Ti-Al oxide surface layer. A detailed analysis of the cyclic voltammograms is discussed in a previous study [60]. The process of oxidation is detected for both substrates. The change of the surface composition during the CV was subsequently analyzed by XPS. The respective spectra are shown in the supporting information, **Figure S4**. The XPS surface composition of both Si | (Ti,Al)N and PC | (Ti,Al)N indicates an increase in the oxygen content after the CV. The oxygen

concentration increases for Si | (Ti,Al)N from 26 at.% to 44 at.% and in the case of PC | (Ti,Al)N from 26 at.% to 47 at.%. In both cases the $O^{2-}$ peak increases significantly, which indicates a $Ti_xAl_yO_z$ layer formation. The nitrogen content decreases in both cases significantly below 1 at.% after the CV measurements. Additionally, a peak shift to higher values was detected in both cases, which indicates the formation of $N_2$. Furthermore, the Ti 2p single spectra show in both cases that the TiN and TiAlON components disappear in the surface-near region and only the $TiO_2$ peak could be detected after the potential cycling. Additionally, the aluminum content decreases, and the titanium content increases in the surface-near region upon CV treatment. Based on these results, the formation of a Ti-rich $Ti_xAl_yO_z$ passivation layer after CV can be concluded.

The EIS measurements were done before and immediately after the CV measurements. The Bode-plots are presented in **Figure 12**.

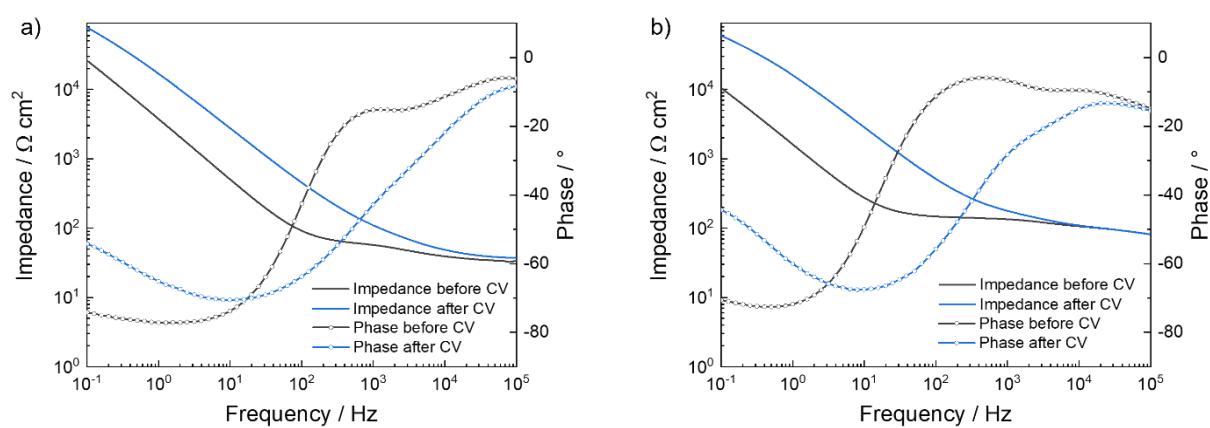

**Figure 12.** EIS measurements in borate buffer (pH 8.3) on a) Si | TiAlN, and b) PC | (Ti,Al)N (before CV after surface oxidation by 15 cycles of CV).

The EIS measurements for the Si | (Ti,Al)N sample are presented in **Figure 12** a showing an increase of the impedance with decreasing frequency. After performing the cyclic voltammograms, the impedance increased due to the formation of a surface

oxide layer. This layer inhibits electrochemical reactions at the electrolyte/coating interface [60]. In **Figure 12** b, the EIS measurements for PC | (Ti,Al)N are shown. The observed higher impedance values at higher frequencies observed for the (Ti,Al)N coating deposited on PC might be due to a reduced electrical conductivity in comparison to the Si | (Ti,Al)N sample.

## 4. Conclusions

The considerably different material properties of polymer substrates and protective hard coatings present a profound challenge for the development of a large-area deposition process ensuring an adhering interface. Here, this challenge was addressed for (Ti,Al)N coatings deposited onto PC by systematically varying the pulsing parameters to reduce the residual stress at the substrate | coating interface. The optimized process with a target peak power density of 0.036 kW cm$^{-2}$ and a duty cycle of 5.3% is categorized in the pulsed DCMS range, whereas all attempted HPPMS processes led to failure of the coating due to cracking and buckling. STEM cross-sections of the adhering (Ti,Al)N coating showed a partly porous microstructure causing residual tensile stress of 2.2 ± 0.2 GPa. The interfacial bond analysis of this PC | (Ti,Al)N sample revealed mainly interfacial CN$_x$ and a smaller fraction of (C-O)-(Ti,Al) bonds, leading to adhesion as verified by the cross-cut test. To evaluate the protective properties of the (Ti,Al)N coating, nanoindentation and EIS measurements were performed revealing an elastic modulus of 296 ± 18 GPa and the formation of a Ti-Al-O passivation layer during oxidation in a borate buffer solution. This work presents a successful strategy of varying deposition parameters to systematically reduce the coating's residual stress caused by the formidable material differences joining at the PC | (Ti,Al)N interface.

## 5. Acknowledgments

This research was funded by the German Research Foundation (DFG, SFB-TR 87/3) "Pulsed high power plasmas for the synthesis of nanostructured functional layers".

## 6. Author contributions (CRediT statement)

**Lena Patterer:** Conceptualization, Methodology, Formal analysis, Investigation, Writing – original draft, Visualization. **Sabrina Kollmann:** Conceptualization, Methodology, Formal analysis, Investigation, Writing – original draft, Visualization. **Teresa de los Arcos:** Methodology, Investigation, Writing – review & editing. **Leonie Jende:** Formal analysis, Investigation, Writing – review & editing. **Soheil Karimi Aghda:** Formal analysis, Investigation, Writing – review & editing. **Damian M. Holzapfel:** Formal analysis, Investigation, Writing – review & editing. **Sameer Salman:** Formal analysis, Investigation, Writing – review & editing. **Stanislav Mráz:** Methodology, Investigation, Writing – review & editing. **Guido Grundmeier:** Conceptualization, Methodology, Writing – original draft, Supervision, Project administration, Funding acquisition. **Jochen M. Schneider:** Conceptualization, Methodology, Writing – original draft, Supervision, Project administration, Funding acquisition.

## 7. Supporting information

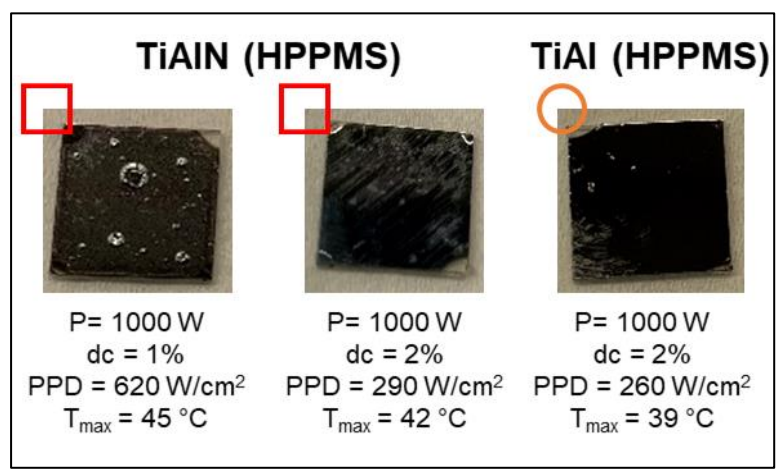

**Figure S1.** Failed HPPMS-(Ti,Al)N and TiAl coatings and their deposition conditions (P = time-averaged power, PPD = target peak power density, dc = duty cycle).

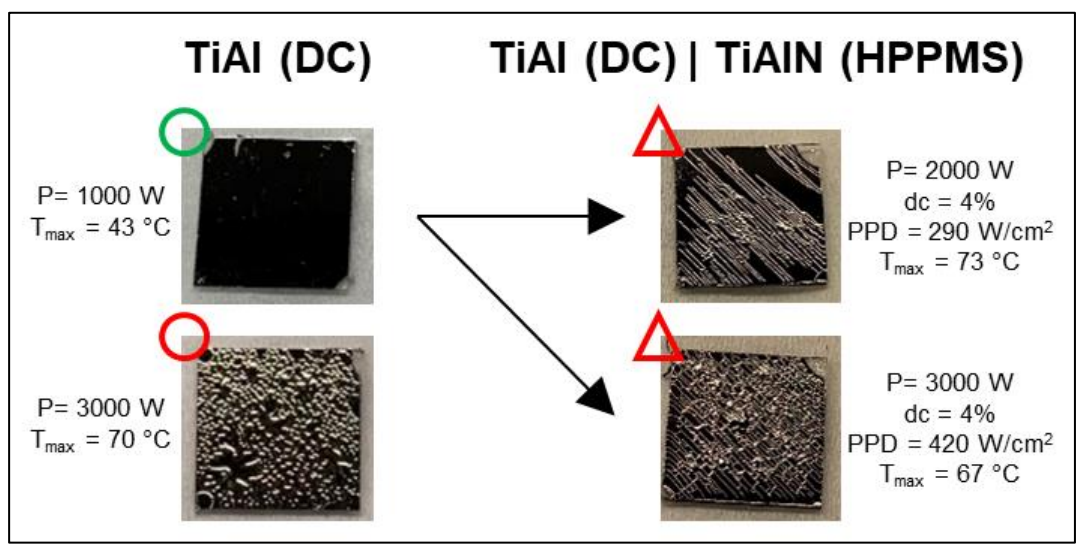

**Figure S2.** DCMS-TiAl coatings and TiAl (DCMS) | (Ti,Al)N (HPPM)-bilayer coatings and their deposition conditions (P = time-averaged power, PPD = target peak power density, dc = duty cycle).

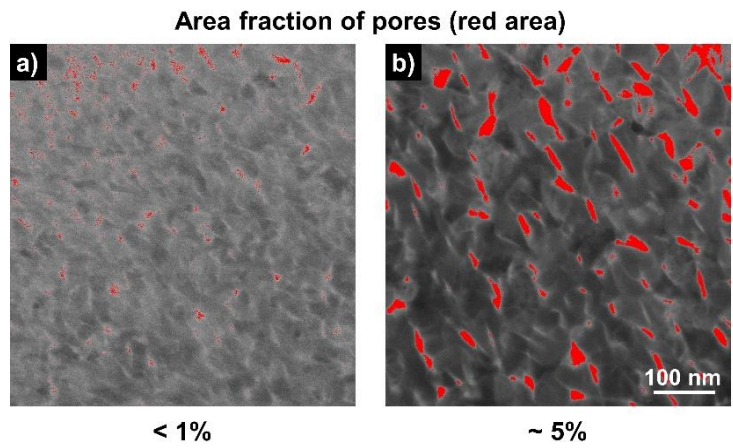

**Figure S3.** Red-marked area fraction of pores at a) the substrate-near region and b) the surface-near region of the diagonal (Ti,Al)N | PC lamella. The pore fraction was analyzed by using the ImageJ software.

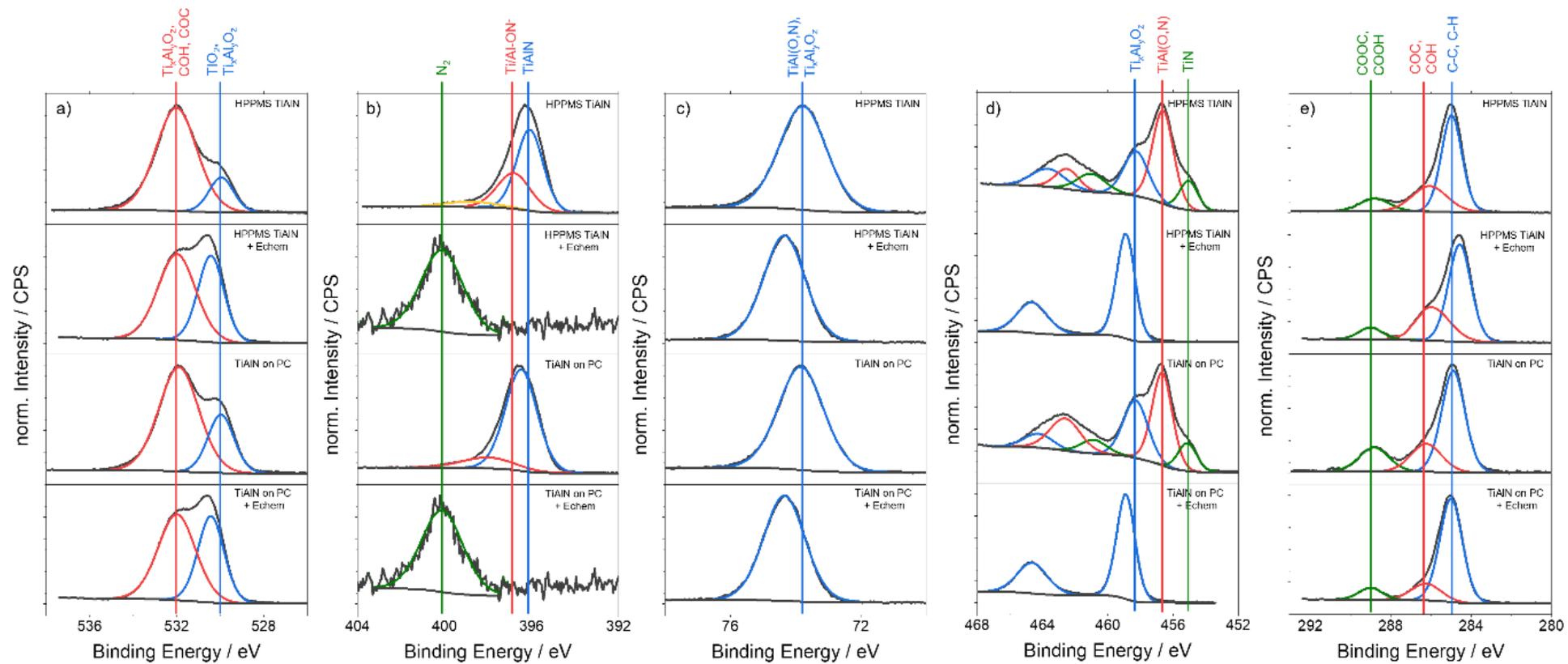

**Figure S4.** XPS a) O 1s, b) N 1s, c) Al 2p, d) Ti 2p, and e) C 1s spectra before and after CV.

# 8. References

[1] GlobalData Plc 2022, Polycarbonate Industry Installed Capacity and Capital Expenditure (CapEx) Forecast by Region and Countries including details of All Active Plants, Planned and Announced Projects, 2021-2025, 2022. https://www.globaldata.com/store/report/polycarbonate-market-analysis/ (accessed 7 November 2022).

[2] J.G. Kim, Chemical recycling of poly(bisphenol A carbonate), Polym. Chem. 11 (2020) 4830–4849. https://doi.org/10.1039/c9py01927h.

[3] A. Kausar, A review of filled and pristine polycarbonate blends and their applications, Journal of Plastic Film & Sheeting 34 (2018) 60–97. https://doi.org/10.1177/8756087917691088.

[4] Y. Zhao, J. Zhang, Q. Xu, H.-Y. Mi, Y. Zhang, T. Li, H. Sun, J. Han, C. Liu, C. Shen, Ultrastable and Durable Silicone Coating on Polycarbonate Surface Realized by Nanoscale Interfacial Engineering, ACS Appl. Mater. Interfaces 12 (2020) 13296–13304. https://doi.org/10.1021/acsami.9b22877.

[5] B. Mailhot, A. Rivaton, J.-L. Gardette, A. Moustaghfir, E. Tomasella, M. Jacquet, X.-G. Ma, K. Komvopoulos, Enhancement of the photoprotection and nanomechanical properties of polycarbonate by deposition of thin ceramic coatings, J. Appl. Phys. 99 (2006) 104310. https://doi.org/10.1063/1.2197030.

[6] S. An, G.-H. Lee, S.W. Jang, Y.-W. Kim, S.H. Lim, D. Kim, S. Han, Al-Si-N/a-SiN:H thin-film coating for polycarbonate glazing applications, Met. Mater. Int. 22 (2016) 535–543. https://doi.org/10.1007/s12540-016-5492-0.

[7] M. to Baben, M. Hans, D. Primetzhofer, S. Evertz, H. Ruess, J.M. Schneider, Unprecedented thermal stability of inherently metastable titanium aluminum nitride by point defect engineering, Materials Research Letters 5 (2017) 158–169. https://doi.org/10.1080/21663831.2016.1233914.

[8] D. McIntyre, J.E. Greene, G. Håkansson, J.-E. Sundgren, W.-D. Münz, Oxidation of metastable single-phase polycrystalline Ti 0.5 Al 0.5 N films: Kinetics and mechanisms, Journal of Applied Physics 67 (1990) 1542–1553. https://doi.org/10.1063/1.345664.

[9] L. Aihua, D. Jianxin, C. Haibing, C. Yangyang, Z. Jun, Friction and wear properties of TiN, TiAlN, AlTiN and CrAlN PVD nitride coatings, International Journal of Refractory Metals and Hard Materials 31 (2012) 82–88. https://doi.org/10.1016/j.ijrmhm.2011.09.010.

[10] M. Lattemann, U. Helmersson, J.E. Greene, Fully dense, non-faceted 111-textured high power impulse magnetron sputtering TiN films grown in the absence of substrate heating and bias, Thin Solid Films 518 (2010) 5978–5980. https://doi.org/10.1016/j.tsf.2010.05.064.

[11] Covestro AG Polycarbonates Business Unit, Makrolon® 2408: ISO datasheet, 2022. https://solutions.covestro.com/en/products/makrolon/makrolon-2408_00013320-05123200?SelectedCountry=DE (accessed 23 May 2023).

[12] S. Karimi Aghda, D.M. Holzapfel, D. Music, Y. Unutulmazsoy, S. Mráz, D. Bogdanovski, G. Fidanboy, M. Hans, D. Primetzhofer, A.S.J. Méndez, A. Anders, J.M. Schneider, Ion kinetic energy- and ion flux-dependent mechanical properties

and thermal stability of (Ti,Al)N thin films, Acta Materialia 250 (2023) 118864. https://doi.org/10.1016/j.actamat.2023.118864.

[13] S. Karimi Aghda, D. Music, Y. Unutulmazsoy, H.H. Sua, S. Mráz, M. Hans, D. Primetzhofer, A. Anders, J.M. Schneider, Unravelling the ion-energy-dependent structure evolution and its implications for the elastic properties of (V,Al)N thin films, Acta Materialia 214 (2021) 117003. https://doi.org/10.1016/j.actamat.2021.117003.

[14] A. Anders, Tutorial: Reactive high power impulse magnetron sputtering (R-HiPIMS), J. Appl. Phys. 121 (2017) 171101. https://doi.org/10.1063/1.4978350.

[15] V. Kouznetsov, K. Macák, J.M. Schneider, U. Helmersson, I. Petrov, A novel pulsed magnetron sputter technique utilizing very high target power densities, Surface and Coatings Technology 122 (1999) 290–293. https://doi.org/10.1016/S0257-8972(99)00292-3.

[16] T. Strunskus, M. Kiene, R. Willecke, A. Thran, C.v. Bechtolsheim, F. Faupel, Chemistry, diffusion and cluster formation at metal-polymer interfaces, Materials and Corrosion 49 (1998) 180–188. https://doi.org/10.1002/(SICI)1521-4176(199803)49:3<180:AID-MACO180>3.0.CO;2-L.

[17] M. Hans, L. Patterer, D. Music, D.M. Holzapfel, S. Evertz, V. Schnabel, B. Stelzer, D. Primetzhofer, B. Völker, B. Widrig, A.O. Eriksson, J. Ramm, M. Arndt, H. Rudigier, J.M. Schneider, Stress-Dependent Elasticity of TiAlN Coatings, Coatings 9 (2019) 24. https://doi.org/10.3390/coatings9010024.

[18] M. Birkholz, Thin film analysis by X-Ray scattering, secondnd reprint, WILEY-VCH, Weinheim, 2009.

[19] N. Bradley, J. Hora, C. Hall, D. Evans, P. Murphy, E. Charrault, Influence of post-deposition moisture uptake in polycarbonate on thin film's residual stress short term evolution, Surface and Coatings Technology 294 (2016) 210–214. https://doi.org/10.1016/j.surfcoat.2016.03.092.

[20] R. Boijoux, G. Parry, C. Coupeau, Buckle depression as a signature of Young's modulus mismatch between a film and its substrate, Thin Solid Films 645 (2018) 379–382. https://doi.org/10.1016/j.tsf.2017.11.011.

[21] K.S. Lee, Effect of elastic modulus mismatch on the contact crack initiation in hard ceramic coating layer, KSME International Journal 17 (2003) 1928–1937. https://doi.org/10.1007/BF02982432.

[22] C. Chaiwong, D.R. McKenzie, M.M.M. Bilek, Cracking of titanium nitride films grown on polycarbonate, Surface and Coatings Technology 201 (2007) 5596–5600. https://doi.org/10.1016/j.surfcoat.2006.07.200.

[23] C. Chaiwong, D.R. McKenzie, M.M.M. Bilek, Study of adhesion of TiN grown on a polymer substrate, Surface and Coatings Technology 201 (2007) 6742–6744. https://doi.org/10.1016/j.surfcoat.2006.09.046.

[24] C. Maurer, U. Schulz, Erosion resistant titanium based PVD coatings on CFRP, Wear 302 (2013) 937–945. https://doi.org/10.1016/j.wear.2013.01.045.

[25] D. Zhang, X. Zuo, Z. Wang, H. Li, R. Chen, A. Wang, P. Ke, Comparative study on protective properties of CrN coatings on the ABS substrate by DCMS and HiPIMS techniques, Surface and Coatings Technology 394 (2020) 125890. https://doi.org/10.1016/j.surfcoat.2020.125890.

[26] P. Pedrosa, M.S. Rodrigues, M.A. Neto, F.J. Oliveira, R.F. Silva, J. Borges, M. Amaral, A. Ferreira, L.H. Godinho, S. Carvalho, F. Vaz, Properties of CrN thin films deposited in plasma-activated ABS by reactive magnetron sputtering, Surface and Coatings Technology 349 (2018) 858–866. https://doi.org/10.1016/j.surfcoat.2018.06.072.

[27] S. Gümüş, Ş. Polat, J.M. Lackner, W. Waldhauser, The Correlation Between Elastic Properties and AFM Images of Nanocoatings on Polymers, Acta Phys. Pol. A 127 (2015) 1142–1144. https://doi.org/10.12693/APhysPolA.127.1142.

[28] M. Kot, Contact mechanics of coating-substrate systems: Monolayer and multilayer coatings, Archives of Civil and Mechanical Engineering 12 (2012) 464–470. https://doi.org/10.1016/j.acme.2012.07.004.

[29] G. Greczynski, L. Hultman, A step-by-step guide to perform x-ray photoelectron spectroscopy, J. Appl. Phys. 132 (2022) 11101. https://doi.org/10.1063/5.0086359.

[30] D.M. Holzapfel, D. Music, S. Mráz, S.K. Aghda, M. Etter, P. Ondračka, M. Hans, D. Bogdanovski, S. Evertz, L. Patterer, P. Schmidt, A. Schökel, A.O. Eriksson, M. Arndt, D. Primetzhofer, J.M. Schneider, Influence of ion irradiation-induced defects on phase formation and thermal stability of Ti0.27Al0.21N0.52 coatings, Acta Materialia (2022) 118160. https://doi.org/10.1016/j.actamat.2022.118160.

[31] M.C. Burrell, J.J. Chera, Polycarbonate Spin Cast Films by XPS, Surface Science Spectra 6 (1999) 1–4. https://doi.org/10.1116/1.1247900.

[32] D.A. Shirley, High-Resolution X-Ray Photoemission Spectrum of the Valence Bands of Gold, Phys. Rev. B 5 (1972) 4709–4714. https://doi.org/10.1103/physrevb.5.4709.

[33] Kratos Analytical Ltd., Relative sensitivity factors (RSF) for Kratos AXIS Supra.

[34] W.C. Oliver, G.M. Pharr, An improved technique for determining hardness and elastic modulus using load and displacement sensing indentation experiments, J. Mater. Res. 7 (1992) 1564–1583. https://doi.org/10.1557/JMR.1992.1564.

[35] M. to Baben, L. Raumann, D. Music, J.M. Schneider, Origin of the nitrogen over- and understoichiometry in Ti(0.5)Al(0.5)N thin films, J. Phys. Condens. Matter 24 (2012) 155401. https://doi.org/10.1088/0953-8984/24/15/155401.

[36] J.T. Gudmundsson, N. Brenning, D. Lundin, U. Helmersson, High power impulse magnetron sputtering discharge, Journal of Vacuum Science & Technology A: Vacuum, Surfaces, and Films 30 (2012) 30801. https://doi.org/10.1116/1.3691832.

[37] J. Wang, Y. Lu, X. Shao, First-Principles Calculation for the Influence of C and O on the Mechanical Properties of γ-TiAl Alloy at High Temperature, Metals 9 (2019) 262. https://doi.org/10.3390/met9020262.

[38] P.H. Mayrhofer, D. Music, J.M. Schneider, Influence of the Al distribution on the structure, elastic properties, and phase stability of supersaturated Ti1−xAlxN, Journal of Applied Physics 100 (2006) 94906. https://doi.org/10.1063/1.2360778.

[39] S. Logothetidis, E.I. Meletis, G. Stergioudis, A.A. Adjaottor, Room temperature oxidation behavior of TiN thin films, Thin Solid Films 338 (1999) 304–313. https://doi.org/10.1016/S0040-6090(98)00975-4.

[40] P. Pokorný, J. Musil, P. Fitl, M. Novotný, J. Lančok, J. Bulíř, Contamination of Magnetron Sputtered Metallic Films by Oxygen From Residual Atmosphere in Deposition Chamber, Plasma Process. Polym. 12 (2015) 416–421. https://doi.org/10.1002/ppap.201400172.

[41] A.Y. Mikheev, V.A. Kharlamov, S.D. Kruchek, A.A. Cherniatina, I.I. Khomenko, Analyzing the contents of residual and plasma-supporting gases inside a vacuum deposition unit chamber, IOP Conf. Ser.: Mater. Sci. Eng. 70 (2015) 12001. https://doi.org/10.1088/1757-899X/70/1/012001.

[42] M.A. Pedreño-Rojas, M.J. Morales-Conde, F. Pérez-Gálvez, P. Rubio-de-Hita, Reuse of CD and DVD Wastes as Reinforcement in Gypsum Plaster Plates, Materials (Basel) 13 (2020). https://doi.org/10.3390/ma13040989.

[43] F. Adibi, I. Petrov, J.E. Greene, L. Hultman, J.-E. Sundgren, Effects of high-flux low-energy (20–100 eV) ion irradiation during deposition on the microstructure and preferred orientation of Ti 0.5 Al 0.5 N alloys grown by ultra-high-vacuum reactive magnetron sputtering, Journal of Applied Physics 73 (1993) 8580–8589. https://doi.org/10.1063/1.353388.

[44] A. Forslund, A. Ruban, Surface energetics of AlxTi1-xN alloys, Computational Materials Science 183 (2020) 109813. https://doi.org/10.1016/j.commatsci.2020.109813.

[45] I. Petrov, P.B. Barna, L. Hultman, J.E. Greene, Microstructural evolution during film growth, Journal of Vacuum Science & Technology A: Vacuum, Surfaces, and Films 21 (2003) S117-S128. https://doi.org/10.1116/1.1601610.

[46] P.B. Barna, M. Adamik, Fundamental structure forming phenomena of polycrystalline films and the structure zone models, Thin Solid Films 317 (1998) 27–33. https://doi.org/10.1016/S0040-6090(97)00503-8.

[47] A. Anders, A structure zone diagram including plasma-based deposition and ion etching, Thin Solid Films 518 (2010) 4087–4090. https://doi.org/10.1016/j.tsf.2009.10.145.

[48] D. Gall, I. Petrov, N. Hellgren, L. Hultman, J.E. Sundgren, J.E. Greene, Growth of poly- and single-crystal ScN on MgO(001): Role of low-energy N2+ irradiation in determining texture, microstructure evolution, and mechanical properties, J. Appl. Phys. 84 (1998) 6034–6041. https://doi.org/10.1063/1.368913.

[49] Y.B. Gerbig, V. Spassov, A. Savan, D.G. Chetwynd, Topographical evolution of sputtered chromium nitride thin films, Thin Solid Films 515 (2007) 2903–2920. https://doi.org/10.1016/j.tsf.2006.08.031.

[50] R. Qiu, O. Bäcke, D. Stiens, W. Janssen, J. Kümmel, T. Manns, H.-O. Andrén, M. Halvarsson, CVD TiAlN coatings with tunable nanolamella architectures, Surface and Coatings Technology 413 (2021) 127076. https://doi.org/10.1016/j.surfcoat.2021.127076.

[51] M. Wiesing, T. de los Arcos, G. Grundmeier, The Thermal Oxidation of TiAlN High Power Pulsed Magnetron Sputtering Hard Coatings as Revealed by Combined Ion and Electron Spectroscopy, Adv. Mater. Interfaces 4 (2017) 1600861. https://doi.org/10.1002/admi.201600861.

[52] H.J. Kim, I.-S. Bae, S.-J. Cho, J.-H. Boo, B.-C. Lee, J. Heo, I. Chung, B. Hong, Synthesis and characteristics of NH2-functionalized polymer films to align and immobilize DNA molecules, Nanoscale Res. Lett. 7 (2012) 30. https://doi.org/10.1186/1556-276X-7-30.

[53] L. Patterer, P. Ondračka, D. Bogdanovski, L. Jende, S. Prünte, S. Mráz, S. Karimi Aghda, B. Stelzer, M. Momma, J.M. Schneider, Bond formation at polycarbonate | X interfaces (X = Ti, Al, TiAl) probed by X-ray photoelectron spectroscopy and density

functional theory molecular dynamics simulations, Applied Surface Science 593 (2022) 153363. https://doi.org/10.1016/j.apsusc.2022.153363.

[54] H. Watanabe, N. Yamada, M. Okaji, Linear Thermal Expansion Coefficient of Silicon from 293 to 1000 K, International Journal of Thermophysics 25 (2004) 221–236. https://doi.org/10.1023/B:IJOT.0000022336.83719.43.

[55] R. Daniel, D. Holec, M. Bartosik, J. Keckes, C. Mitterer, Size effect of thermal expansion and thermal/intrinsic stresses in nanostructured thin films: Experiment and model, Acta Materialia 59 (2011) 6631–6645. https://doi.org/10.1016/j.actamat.2011.07.018.

[56] H. Rueß, D. Music, A. Bahr, J.M. Schneider, Effect of chemical composition, defect structure, and stress state on the elastic properties of (V$_{1-x}$Al$_x$)$_{1-y}$N$_y$, J. Phys. Condens. Matter 32 (2020) 25901. https://doi.org/10.1088/1361-648X/ab46df.

[57] A.R. Boccaccini, G. Ondracek, P. Mazilu, D. Windelberg, On the Effective Young's Modulus of Elasticity for Porous Materials: Microstructure Modelling and Comparison Between Calculated and Experimental Values, Journal of the Mechanical Behavior of Materials 4 (1993) 119–128. https://doi.org/10.1515/JMBM.1993.4.2.119.

[58] M. Hans, M. to Baben, D. Music, J. Ebenhöch, D. Primetzhofer, D. Kurapov, M. Arndt, H. Rudigier, J.M. Schneider, Effect of oxygen incorporation on the structure and elasticity of Ti-Al-O-N coatings synthesized by cathodic arc and high power pulsed magnetron sputtering, Journal of Applied Physics 116 (2014) 93515. https://doi.org/10.1063/1.4894776.

[59] T.-H. Fang, W.-J. Chang, Nanoindentation characteristics on polycarbonate polymer film, Microelectronics Journal 35 (2004) 595–599. https://doi.org/10.1016/j.mejo.2004.02.004.

[60] M. Wiesing, M. to Baben, J.M. Schneider, T. de los Arcos, G. Grundmeier, Combined Electrochemical and Electron Spectroscopic Investigations of the Surface Oxidation of TiAlN HPPMS Hard Coatings, Electrochimica Acta 208 (2016) 120–128. https://doi.org/10.1016/j.electacta.2016.05.011.